\documentclass{article}
\usepackage{graphicx} 
\usepackage{booktabs}
\usepackage{tabularx}
\usepackage[a4paper, margin=2.5cm]{geometry}
\usepackage{multirow}

\title{Benchmarking Recent Single-Cell Synthetic Data Generation Methods on the OneK1K Dataset}

\author{
  Buse Giledereli\thanks{These authors contributed equally (co-first authors).}\\
  Computer Engineering\\
  Bo\u{g}azi\c{c}i University, Istanbul, Turkey\\
  \texttt{buse.giledereli@bogazici.edu.tr}
  \and
  Ba\c{s}ar Temiz\footnotemark[1]\\
  Computer Engineering\\
  Bo\u{g}azi\c{c}i University, Istanbul, Turkey\\
  \texttt{basar.temiz@std.bogazici.edu.tr}
  \and
  Amirreza Sattarzadeh Khanehbargh\footnotemark[1]\\
  Computer Engineering\\
  Bo\u{g}azi\c{c}i University, Istanbul, Turkey\\
  \texttt{amirreza.sattarzadeh@std.bogazici.edu.tr}
}

\date{} 

\begin{document}

\maketitle

\begin{abstract}
Synthetic single-cell RNA-seq data generation offers a promising route for privacy-aware biomedical data sharing, but donor-linked datasets require careful balancing of biological utility and privacy risk. In this work, we benchmark synthetic generation methods for the CAMDA 2026 ELSA Health Privacy Challenge using the donor-split OneK1K dataset, comparing scGFT, scLDM, and Cell2Sentence across statistical fidelity, downstream biological preservation, and membership inference vulnerability. Because full-transcriptome generation is computationally demanding, we focus generative modeling on highly variable genes and evaluate alternative strategies for reconstructing non-HVG expression. Our results show that non-HVG handling is a central determinant of privacy: directly copying non-HVG values leads to severe leakage, whereas Poisson-based reconstruction substantially reduces membership inference risk. We further find that scGFT and scLDM better preserve biological structure than an independent Poisson baseline, while maintaining near-random privacy profiles under appropriate configurations. These findings highlight the importance of evaluating synthetic single-cell data beyond marginal similarity alone, emphasizing robustness, interpretability, and privacy-conscious design choices throughout preprocessing, generation, and post-processing.
\end{abstract}

\section{Introduction}
 
Single-cell RNA sequencing has transformed biomedical research by enabling the characterization of cellular heterogeneity at high resolution. However, these single-cell RNA sequencing datasets are high-dimensional, costly to generate, donor-linked, and may contain sensitive biological information. Synthetic single-cell data generation has therefore emerged as a promising strategy for enabling privacy-aware data sharing, augmenting scarce datasets, and supporting the pretraining or evaluation of increasingly large single-cell models \cite{scgpt, cscgan, scdesign3}. In this setting, a useful synthetic dataset should preserve biologically meaningful structure while limiting the leakage of donor-specific information.

This requirement is particularly important for donor-linked cohorts, where synthetic samples may inadvertently reveal whether a donor or cell profile was present in the training data. Membership inference and related privacy attacks have shown that synthetic biomedical data can remain vulnerable even when direct identifiers are removed \cite{mia-synthetic-health-data}. Thus, the generative models must be assessed not only by their fidelity to real expression distributions, but also by their robustness against privacy risks.

The CAMDA 2026 ELSA Health Privacy Challenge provides a benchmark setting for studying this trade-off, by using a donor-split OneK1K dataset \cite{onek1k} with 633,711 training cells from 490 donors and 634,022 test cells from 491 donors across 25,834 genes. We investigate this trade-off using a broad range of generative modeling approaches: Single-cell Graph Fourier Transform (scGFT) \cite{scgft} as a train-free, Fourier-based generator, Single-cell Latent Diffusion Model (scLDM) \cite{scldm} as a diffusion-based model, and Cell2Sentence \cite{cell2sentence} as a language model–driven method, and assess their performance in terms of fidelity, biological plausibility, and privacy.

Our study is motivated by a practical engineering challenge: subject to constraints on memory, computational runtime, and reproducibility, we seek to determine which modeling choices enhance generative performance while explicitly accounting for the privacy–utility trade-off. We explored this through iterative implementation and analysis across various method families. scGFT gives direct control over diversity by perturbing Fourier components, while scLDM offers strong latent generative capacity on informative genes but requires careful post-processing when the working feature space is reduced. Sequence-based approaches such as Cell2Sentence provide interpretability but face scale and computational cost challenges.  


We present two primary contributions. First, we provide a privacy-aware benchmark comparison of multiple synthetic single-cell generation paradigms, evaluating spectral, diffusion-based, and sequence-based methods not only by utility but also by donor-level privacy risk. Second, we investigate practical solutions for handling non-HVG genes, showing that decisions made outside the primary modeling space can substantially affect both full-transcriptome utility and privacy. Together, this synthetic single-cell benchmarking framework enables us to evaluate models simultaneously in terms of biological plausibility, behavior in feature space, and privacy-aware data generation.

\section{Methods}

In our benchmarking study for the single-cell gene expression, we evaluated the fidelity and privacy of three distinct generative architectures: Single-cell Graph Fourier Transform (scGFT), Single-cell Latent Diffusion Models (scLDM) and Cell2Sentence on OneK1K dataset \cite{onek1k}.  The OneK1K dataset is a large-scale resource for single-cell transcriptomic analyses, specifically developed to investigate the genetic determinants underlying inter-individual variation in gene expression within human immune cell populations. It consists of single-cell RNA sequencing (scRNA-seq) data from approximately 1.2 million peripheral blood mononuclear cells (PBMCs) collected from 982 individual donors \cite{onek1k}. 

For our benchmarking analysis, we began with the full feature set of 25,834 genes, representing the broad transcriptomic landscape of the immune system. However, due to the substantial computational demands of generative modeling and the time constraints of the challenge, we restricted the modeling phase to highly variable genes (HVGs), which capture the main axes of biological variation. This choice is also consistent with common practice in single-cell generative modeling, where HVG selection is often used to reduce dimensionality while preserving the most informative biological signals. Data preprocessing and HVG selection were performed using the Scanpy library \cite{scanpy}. Privacy evaluation was performed using the GAN-leaks method \cite{gan-leaks}. By focusing the more computationally intensive generative modeling on this informative subset, we were able to evaluate model performance efficiently while remaining aligned with the operational and privacy constraints of the challenge.

\subsection{scLDM}

scLDM (Palla et al., 2025) is a two-stage latent-diffusion generator: a transformer VAE first compresses single-cell expression into a low-dimensional continuous latent, after which a Diffusion Transformer (DiT) is trained to denoise that latent under classifier-free guidance on cell type. Generation samples a latent from the diffusion prior, decodes through the VAE, and reads out per-gene Negative Binomial counts conditioned on per-cell-type log-library-size factors estimated from the train set. We trained scLDM on the top 1{,}000 highly variable genes (HVGs) of OneK1K for $50+50$ epochs (VAE then DiT) with a guidance weight of $1.0$ on \texttt{cell\_type}.

Because scLDM operates on an HVG-1000 subset, the raw output is a $1{,}000$-gene synthetic that does not align with the full $25{,}834$-gene MIA test set. We evaluate two variants: (i) the raw HVG-only output, for which we additionally pre-subset the MIA test AnnData to the same $1{,}000$ HVGs so the attack pipeline operates on a matched gene space, and (ii) a \emph{hybrid} that hstacks the scLDM HVG-1000 counts with per-(\texttt{cell\_type}, gene) Poisson samples for the remaining $24{,}834$ non-HVG genes, drawn from the full-gene train file. The Poisson fill is closed-form ($\lambda_{g,c}$ is the train mean of gene $g$ in cells of type $c$), adds no donor information, and lets the synthetic file plug into the standard pipeline without HVG pre-subsetting.

scLDM's classifier-free guidance rollout writes both halves of every batch: unconditional and conditional. Only the conditional half corresponds to the labelled distribution we evaluate; our wrapper drops the unconditional half so the synthetic AnnData has the expected 53,414 labelled cells rather than 106,828 mixed cells.

\subsection{Single-cell Graph Fourier Transform (scGFT)}

The scGFT framework is a training-free, fully analytical methodology specifically developed to solve the extensive data demands and computational requirements associated with conventional generative modeling paradigms, such as Generative Adversarial Networks (GANs) and Variational Autoencoders (VAEs). Its methodology leverages the Discrete Fourier Transform (DFT) to map individual gene expression profiles into a complex space, where stochastically selected components are modified before being reconstructed via Inverse Fourier Transform (IFT). By utilizing the Fourier Transform, the model shifts from a focus on the overall geometric layout of the entire population to a more localized, cell-centric view. This approach treats gene expression as a graph signal, enabling fine-grained variations that capture individual cell behavior while preserving consistency with the broader system. The hyperparameter controlling how many components are modified governs the trade-off between diversity and biological fidelity: higher values increase deviation from original cells to explore more diverse states. Overall, scGFT is computationally efficient, scalable, and can synthesize entire populations from the expression profile of a single cell.

The original study assessed performance using between 5 and 100 modified components, but did not take privacy into consideration. Our experiments indicate that the extent of modification has a direct effect on both the diversity and privacy of the generated output. When fewer components are altered, the resulting synthetic cells stay mathematically very close to their original counterparts, preserving high fidelity but remaining susceptible to MIA attacks. As the number of modified components increases, the synthetic population becomes more diverse, which reduces vulnerability to attacks but also decreases fidelity.

In its standard implementation, the algorithm focuses primarily on Highly Variable Genes (HVGs) to capture biological variability associated with distinct cellular states. During the synthesis step, the original expression values of non-HVG (invariant) genes are typically retained and reinserted into the synthetic expression matrix to preserve accurate correspondence and alignment between cells. However, leaving these genes unchanged introduces a substantial privacy risk: the presence of identical non-HVG features in both synthetic and training cells enables straightforward inference of the original training data. To mitigate this vulnerability, we developed a modified version of the framework in which non-HVG gene expression is synthesized by sampling from a Poisson distribution parameterized by the corresponding mean expression levels estimated from the training data. This adjustment effectively addresses the security weakness of earlier models, which made the system vulnerable to MIAs. In our implementation, we used the default settings to identify HVGs, yielding 1,118 genes to which we applied DFT, and we sampled the remaining invariant genes from a Poisson distribution.

\subsection{Cell2Sentence}

The Cell2Sentence (C2S) methodology bridges the gap between natural language processing and biology by transforming single-cell transcriptomic data into textual sequences known as cell sentences. The process begins by preprocessing transcript count matrices to form cell sentences, rank-ordering gene names by expression level to convert numerical biological data into a format that LLMs can interpret. To maintain data integrity, C2S utilizes a linear regression model to map the relationship between a gene's log-rank and its original expression, ensuring the transformation is robust and reversible with minimal information loss. These cell sentences are then used to fine-tune LLMs for various tasks, including conditional cell generation, cell-type annotation, and the extraction of natural language biological insights.

Although the model shows promise, the computational cost of full-scale inference, which was estimated at several single GPU-weeks, was prohibitive for this task. Consequently, we limited synthetic data generation to 1\% of the total cells and utilized a fine-tuned iteration provided by the authors. This model, C2S-Pythia-410m-cell-type-conditioned-cell-generation, is built upon the Pythia-410m architecture and fine-tuned using the Cell2Sentence (C2S) framework. The model was trained on a diverse dataset of over 57 million human and mouse cells from the CellxGene and Human Cell Atlas repositories, focusing on the top 200 genes per cell. Its primary utility lies in cell type-conditioned generation, enabling the synthesis of realistic single-cell profiles that adhere to the gene expression patterns of specific, user-defined cell types.

\section{Results}

We evaluate the synthetic datasets along three axes: statistical fidelity, downstream biological preservation, and privacy risk. In terms of fidelity (Table~\ref{tab:fidelity}), the Poisson baseline remains a strong competitor, achieving the highest Spearman correlation for the scGFT comparison (\texttt{cc\_spear} $=0.998$). However, scGFT maintains competitive statistical similarity (\texttt{cc\_spear} $=0.984$) and slightly improves MMD, suggesting that the Fourier-based representation does not substantially sacrifice marginal fidelity. A similar pattern is observed for scLDM, the hybrid variant closely matches the Poisson baseline on per-gene rank correlation (\texttt{cc\_spear} $= 0.995$), while improving Pearson correlation ($0.999$) and per-gene EMD ($0.009$). The scGFT model with 25 components modified performs best on gene-gene correlation preservation (\texttt{corr\_corr} $=0.856$) for full-dataset generation, whereas scLDM performs best for the shared-latent mixing metric (LISI $=0.695$), and both generators performs better than the Poisson baseline with MMD at $10^{-4}$. Even though scCell2SentenceModel was not fine-tuned and generated only 200 genes per cell, its performance remains comparable to scGFT and scLDM, particularly in terms of correlation scores. 

The main advantage of the learned generative approaches becomes clearer in the downstream biological evaluation (Table~\ref{tab:downstream}). Although the Poisson baseline can reproduce simple marginal and cluster-level statistics, it performs poorly when real and synthetic cells are jointly clustered. In contrast, both scGFT and scLDM better preserve the underlying biological manifold. This is reflected in the GT vs.~Comb.\ ARI metric, where scGFT (0.342) substantially outperforms the Poisson baseline (0.220) , and where scLDM achieves ARI values of $0.462$ for the HVG-only variant and $0.423$ for the hybrid variant. Even the off-the-shelf scCell2Sentence model achieves higher scores than the baseline Poisson model for GT vs Comb. These results indicate that the learned models capture dependencies between genes and cell states that are lost under independent Poisson sampling, allowing synthetic cells to retain their functional identity and integrate more consistently with real populations. Also we note that the performance drop of scLDM on the Real vs Syn ARI metric is due to the model’s random generation order. In this evaluation, the generated data is expected to preserve the same ordering as the original data; when this alignment is not maintained, it results in a low performance.

Privacy remains an important constraint in selecting a usable synthetic-data generator (Table~\ref{tab:privacy}). For scGFT, the copied non-HVG variant is clearly non-viable, reaching perfect membership inference leakage (MIA AUC $=1.0$). The lower-complexity Fourier configuration ($n=10$) is also more vulnerable than the Poisson baseline. However, increasing the number of Fourier components to $n=25$ reduces leakage to a near-random privacy profile (AUC $=0.494$), comparable to the Poisson baseline. This suggests that scGFT can achieve stronger biological utility without a privacy penalty when its synthesis complexity is properly tuned. Similarly, scLDM attains an AUC of 0.500, indicating a random distribution and demonstrating that both models successfully meet the privacy constraints.

\begin{table}[htbp]
\centering
\caption{Synthetic Data Fidelity \& Statistical Metrics}
\begin{tabularx}{\linewidth}{ll *{6}{>{\centering\arraybackslash}X}}
\toprule
\textbf{Model} & \textbf{Config} & \textbf{cc\_spear} & \textbf{cc\_pear} & \textbf{mmd} & \textbf{emd} & \textbf{corr\_corr} & \textbf{lisi} \\
\midrule
Poisson & Baseline & 0.9983 & 0.9525 & 0.0003 & 0.0201 & 0.7357 & 0.0000 \\
\midrule
\multirow{3}{*}{scGFT} 
& $n=10$, non-HVG poisson & 0.9839 & 0.9473 & 0.0001 & 0.0210 & 0.8556 & 0.0000 \\
& $n=25$, non-HVG poisson & 0.9842 & 0.9473 & 0.0001 & 0.0209 & 0.8560 & 0.0000 \\
& $n=25$, non-HVG copied & 0.9866 & 0.9983 & 0.0001 & 0.0120 & 0.8995 & 0.2656 \\
\midrule
\multirow{2}{*}{scLDM}
& HVG-1000 only           & 0.9906 & 0.9927 & 0.0001 & 0.0703 & 0.8838 & 0.6952 \\
& HVG-1000 + Poisson fill & 0.9952 & 0.9990 & 0.0001 & 0.0092 & 0.8422 & 0.5177 \\
\midrule
scCell2Sentence & 1\% data & 0.9554 & 0.9815 & 0.0003 & 0.0308 & 0.2761 & 0.1930 \\
\bottomrule
\end{tabularx}
\label{tab:fidelity}
\end{table}

\begin{table}[htbp]
\centering
\caption{Clustering and Downstream Performance Metrics}
\label{tab:clustering_metrics}
\begin{tabular}{llcc}
\toprule
\textbf{Model} & \textbf{Config} 
& \multicolumn{2}{c}{\textbf{Clustering ARI}} \\
\cmidrule(lr){3-4}
& 
& \textbf{Real vs Syn} 
& \textbf{GT vs Comb} \\
\midrule
Poisson & Baseline & 0.7413 & 0.2196 \\
\midrule
\multirow{3}{*}{scGFT}
& $n=10$, non-HVG poisson & 0.4383 & 0.3148\\
& $n=25$, non-HVG poisson & 0.4884 & 0.3423 \\
& $n=25$, non-HVG copied & 0.3876 & 0.4645\\
\midrule
\multirow{2}{*}{scLDM}
& HVG-1000 only           & 0.0011 & 0.4615\\
& HVG-1000 + Poisson fill & 0.0006 & 0.4232 \\
\midrule
scCell2Sentence & 1\% data & 0.2535 & 0.3099 \\
\bottomrule
\end{tabular}
\label{tab:downstream}
\end{table}

\begin{table}[htbp]
\centering
\caption{Privacy \& Membership Inference Risk}
\label{tab:privacy_metrics_1}
\begin{tabular}{llrrrr}
\toprule
\textbf{Model} & \textbf{Config} 
& \textbf{AUC-ROC} 
& \textbf{Acc (best)} 
& \textbf{Avg Precision} 
& \textbf{PR-AUC} \\
\midrule
Poisson & dist & 0.5252  & 0.5400 & 0.5636 & 0.5560 \\
\midrule
\multirow{3}{*}{scGFT}
& $n=10$, non-HVG poisson & 0.7392  & 0.7200 & 0.6834 & 0.6747 \\
& $n=25$, non-HVG poisson & 0.4936  & 0.5100 & 0.5211 & 0.5115 \\
& $n=25$, non-HVG copied & 1.0000  & 1.0000 & 1.0000 & 1.0000 \\
\midrule
\multirow{2}{*}{scLDM}
& HVG-1000 only           & 0.5304 & 0.5400 & 0.5286 & 0.5828 \\
& HVG-1000 + Poisson fill & 0.5000 & 0.5100 & 0.5022 & 0.4892 \\
\bottomrule
\end{tabular}
\label{tab:privacy}
\end{table}

\section{Discussion}

In this work, we benchmarked three generative architectures: scGFT, scLDM, and Cell2Sentence on the donor-split OneK1K dataset to study the privacy-utility trade-off in synthetic single-cell RNA-seq generation. To accommodate the computational and time constraints of the challenge, we restricted generative modeling to highly variable genes (HVGs) and reconstructed non-HVGs through either direct copying or stochastic Poisson sampling, allowing us to isolate the effect of background-gene handling on both fidelity and leakage. We assessed each configuration along three complementary axes: statistical fidelity, downstream biological preservation, and membership inference risk. This design lets us examine not only which generator performs best in aggregate, but also how specific modeling choices, Fourier complexity, latent-space generation, HVG versus full-gene treatment, shift the balance between biological utility and donor-level privacy.

Our results highlight several patterns worth emphasizing. The dominant privacy lever in our experiments is the handling of non-HVG genes: the most informative comparison is between two scGFT configurations differing only in this treatment, where copying non-HVGs directly from training cells enables successful membership inference attacks, while stochastic Poisson replacement substantially reduces leakage. Beyond this, Fourier complexity provides a practical control over the privacy utility frontier, where increasing the number of perturbed components improves privacy with limited fidelity cost. Learned generators also capture joint structure that independent sampling cannot the Poisson baseline is a strong competitor on marginal statistics but collapses on joint structure, and both scGFT and scLDM substantially exceed it in this regard. Metric interpretation, however, requires care: the near-zero Real-vs-Syn ARI for scLDM reflects the metric's assumption of preserved cell ordering rather than a fidelity failure, as the order-independent GT-vs-Comb. ARI tells the opposite story, and more generally we observe that metrics are not correlated with one another, so a model can appear reliable on one and drift on another. These observations are further shaped by the dataset itself: OneK1K is a strong benchmark for privacy-preserving synthesis given its scale and donor diversity, but its high sparsity and dimensionality impose computational constraints that motivate reduced representations and HVG-restricted modeling, and our profiling of the test split shows high global sparsity and a long tail of rarely detected genes that make metric behavior sensitive to filtering and representation choices, with HVG subsets capturing only part of the total expression mass. Together, these findings motivate a broader evaluation perspective for Track II that prioritizes robustness and interpretability across multiple metrics, reporting both HVG-focused and full-gene behavior, including transform robustness checks, and interpreting privacy through donor level diagnostics rather than single aggregate metrics alone.

\bibliographystyle{plain}
\bibliography{reference}

\end{document}